\documentclass[prb,preprint,showpacs,floatfix]{revtex4}

\usepackage{graphicx}
\usepackage{tabularx}
\usepackage{array}
\usepackage{amsmath,amsfonts,amssymb}
\usepackage[ansinew]{inputenc}
\usepackage{color}
\usepackage{calc}
\usepackage{braket}
\usepackage{array}
\usepackage{color}
\usepackage{dcolumn}
\usepackage{bm}
\usepackage{mathrsfs}
\usepackage[normalem]{ulem}
\usepackage{booktabs}
\usepackage{multirow}
\usepackage{subfigure}
\usepackage{epsfig}
\usepackage{amsmath,amsfonts,amssymb}
\usepackage{physics}
\usepackage{tikz}

\newcommand{\Figref}[1]{Fig.~\ref{#1}}
\newcommand{\ExFigref}[1]{Extended Data Fig.~\ref{#1}}

\begin{document}
\title{Energy funnelling within multichromophore architectures monitored with subnanometre resolution}

\author
{Shuiyan Cao$^{1,2,\dagger}$, Anna Ros\l awska$^{1,\dagger,\ast}$, Benjamin Doppagne$^1$, Michelangelo Romeo$^1$, Michel F\'eron$^3$, Fr\'ed\'eric Ch\'erioux$^3$, Herv\'e Bulou$^1$,\\
 Fabrice Scheurer$^1$, Guillaume Schull$^{1,\ast}$\\
\normalsize{$^1$ Universit\'e de Strasbourg, CNRS, IPCMS, UMR 7504, F-67000 Strasbourg, France,} \\
\normalsize{$^2$ Department of Applied Physics, Nanjing University of Aeronautics and Astronautics, Nanjing, 210016,  China,} \\
\normalsize{$^3$ Universit\'e Bourgogne Franche-Comt\'e, FEMTO-ST, UFC, CNRS, 15B avenue des Montboucons, F-25030 Besan\c con, France.} \\
\altaffiliation{guillaume.schull@ipcms.unistra.fr}
\altaffiliation{Anna.Roslawska@ipcms.unistra.fr}
}

\footnotetext[2]{These authors contributed equally to this paper.}

\date{\today}

\maketitle

In natural \cite{Mirkovic2017} and artificial \cite{Balzani2008} light-harvesting complexes (LHC) the resonant energy transfer (RET) between chromophores enables an efficient and directional transport of solar energy between collection and reaction centers. The detailed mechanisms involved in this energy funneling are intensely debated \cite{Arsenault2020, Ma2019}, essentially because they  rely on a succession of individual RET steps that can hardly be addressed separately. Here, we developed a scanning tunnelling microscopy-induced luminescence (STML) approach allowing visualizing, addressing and manipulating energy funneling within multi-chromophoric structures with sub-molecular precision. We first rationalize the efficiency of the RET process at the level of chromophore dimers. We then use highly resolved fluorescence microscopy (HRFM) maps \cite{Doppagne2020} to follow energy transfer paths along an artificial trimer of descending excitonic energies which reveals a cascaded RET from high- to low-energy gap molecules. Mimicking strategies developed by photosynthetic systems, this experiment demonstrates that intermediate gap molecules can be used as efficient ancillary units to convey energy between distant donor and acceptor chromophores. Eventually, we demonstrate that the RET between donors and acceptors is enhanced by the insertion of passive molecules acting as non-covalent RET bridges. This mechanism, that occurs in experiments performed in inhomogeneous media and which plays a decisive role in fastening RET in photosynthetic systems \cite{Pullerits1997,Scholes2000}, is reported at the level of individual chromophores with atomic-scale resolution. As it relies on organic chromophores as elementary components, our approach constitutes a powerful model to address fundamental physical processes at play in natural LHC.\\

\begin{figure}
  \includegraphics[width=89 mm]{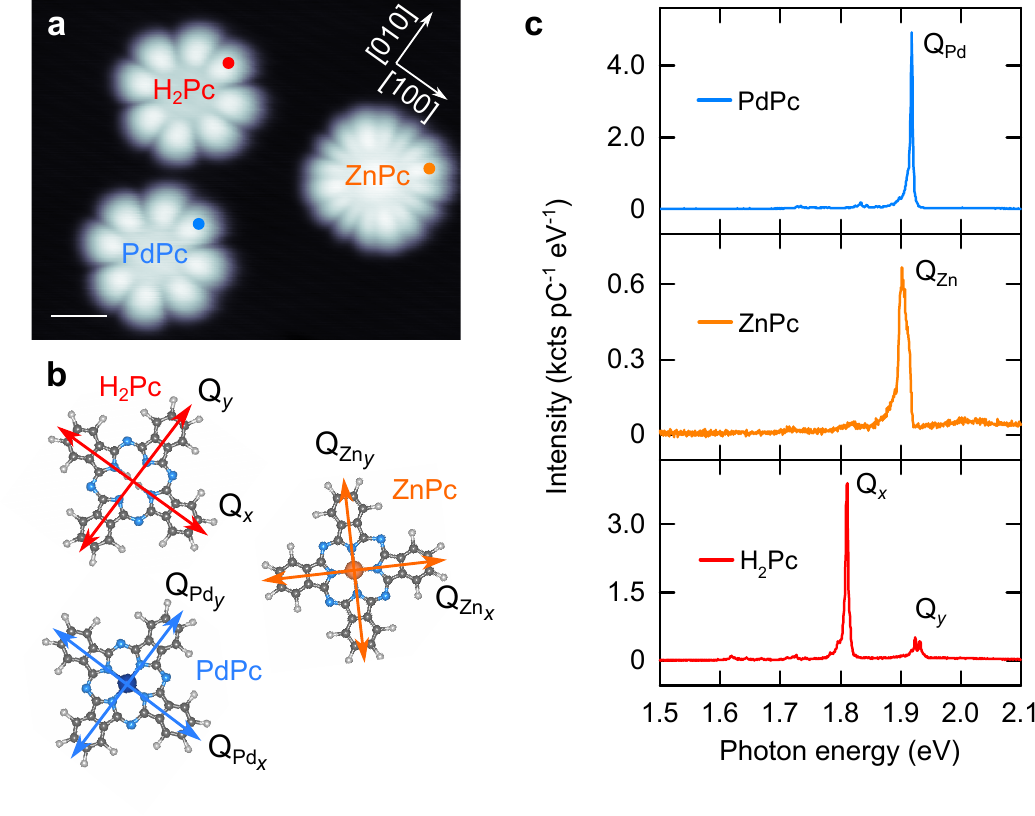}
  \caption{\label{fig1}  \textbf{STM-induced fluorescence of individual chromophores.} (a) STM image ($I$ = 10 pA, $V$ = -2.5 V, scale bar 1 nm), (b) ball-and-stick models, and (c) typical STML spectra ($V=$ -2.5 V , acquisition time $t$ = 60 s, from top to bottom $I$ = 250 pA, 50 pA, 50 pA) of the three chromophores (PdPc, ZnPc and H$_{2}$Pc) investigated in this study. The molecules are separated from the Ag(111) sample by three layers of NaCl. The spectra were recorded with the tip positioned at the extremity of a pyrrole sub-unit (colored dots in (a)). Spectral features in (c) are associated to degenerate (Q$_\textrm{Pd}$, Q$_\textrm{Zn})$ and non-degenerate (Q$_{x}$, Q$_{y}$) radiative transitions of the chromophore dipoles, whose orientations are given in (b).} 
\end{figure}

The energy funneling in LHC relies on cascaded RET events, based primarily on dipole-dipole interactions, occurring between high and low energy-gap chromophores. The efficiency of the process is further improved by exchange interaction mediated energy transfers at short distance \cite{Andrews2011}, delocalization of the excitation over coherently coupled molecules involving vibronic coupling or not\cite{Engel2007}, fine-tuning of the absorption/emission energy of ancillary chromophores \cite{Mirkovic2017}, or promoted energy transfer by the environment \cite{Pullerits1997,Scholes2000}. The influence of these parameters on the extraordinary ability of biological LHC to transfer energy between distant centers remains to be clarified \cite{Arsenault2020,Ma2019} and their systematic control in artificial systems is both crucial and far ahead \cite{Lim2015,Trofymchuk2017,brian2020}. 
Recent experiments\cite{Qiu2003,Merino2015,Chong2016,Zhang2016,Imada2016,Doppagne2017,Imada2017,zhang2017a,Dolezal2019,Doppagne2020} have shown that a cryogenic STM can be used to probe the fluorescent properties of individual or interacting chromophores with high spatial, spectral and temporal resolution. 
Here, we develop this approach to follow the energy funneling within multi-chromophoric architectures and characterize the decisive role played by intermediate molecules, located between donor and acceptor units, in promoting, blocking or directing RET. To this end, we used three chromophores, palladium-phthalocyanine (PdPc), zinc-phthalocyanine (ZnPc) and free-base phthalocyanine (H$_{2}$Pc), deposited on an insulating NaCl trilayer covering Ag(111) (see Methods). STM images of ZnPc recorded at negative voltages (\Figref{fig1}a) reveal 16-lobe structures resulting from the fast motion of ZnPc between two metastable adsorption sites \cite{Zhang2016,Imada2016,Doppagne2017,Patera2019a, Dolezal2019}, whereas PdPc and H$_{2}$Pc exhibit 8-lobe patterns characteristic of the highest occupied molecular orbitals (HOMO) of fixed phthalocyanine molecules.

A straightforward assignment of the molecules is obtained from their STML spectra (\Figref{fig1}c). PdPc is characterized by a sharp emission line (Q$_\textrm{Pd}$) at $\approx$ 1.92 eV. The main emission line in the ZnPc spectrum (Q$_\textrm{Zn}$) is less intense and located at a slightly lower energy ($\approx$ 1.90 eV), attesting for the smaller optical gap of the molecule. For both molecules, the Q-line corresponds to two degenerated transitions associated to dipoles oriented along the main molecular axes (\Figref{fig1}b). H$_{2}$Pc is characterized by one intense emission line at $\approx$ 1.81 eV (Q$_{x}$) and a weak emission at $\approx$ 1.93 eV (Q$_{y}$) corresponding to dipoles oriented respectively along and perpendicular to the inner H-H axis of the molecule. The purely electronic transitions of PdPc, H$_{2}$Pc and ZnPc are accompanied by vibronic lines at lower energies\cite{Doppagne2017}. 

\begin{figure}
  \includegraphics[width=\linewidth]{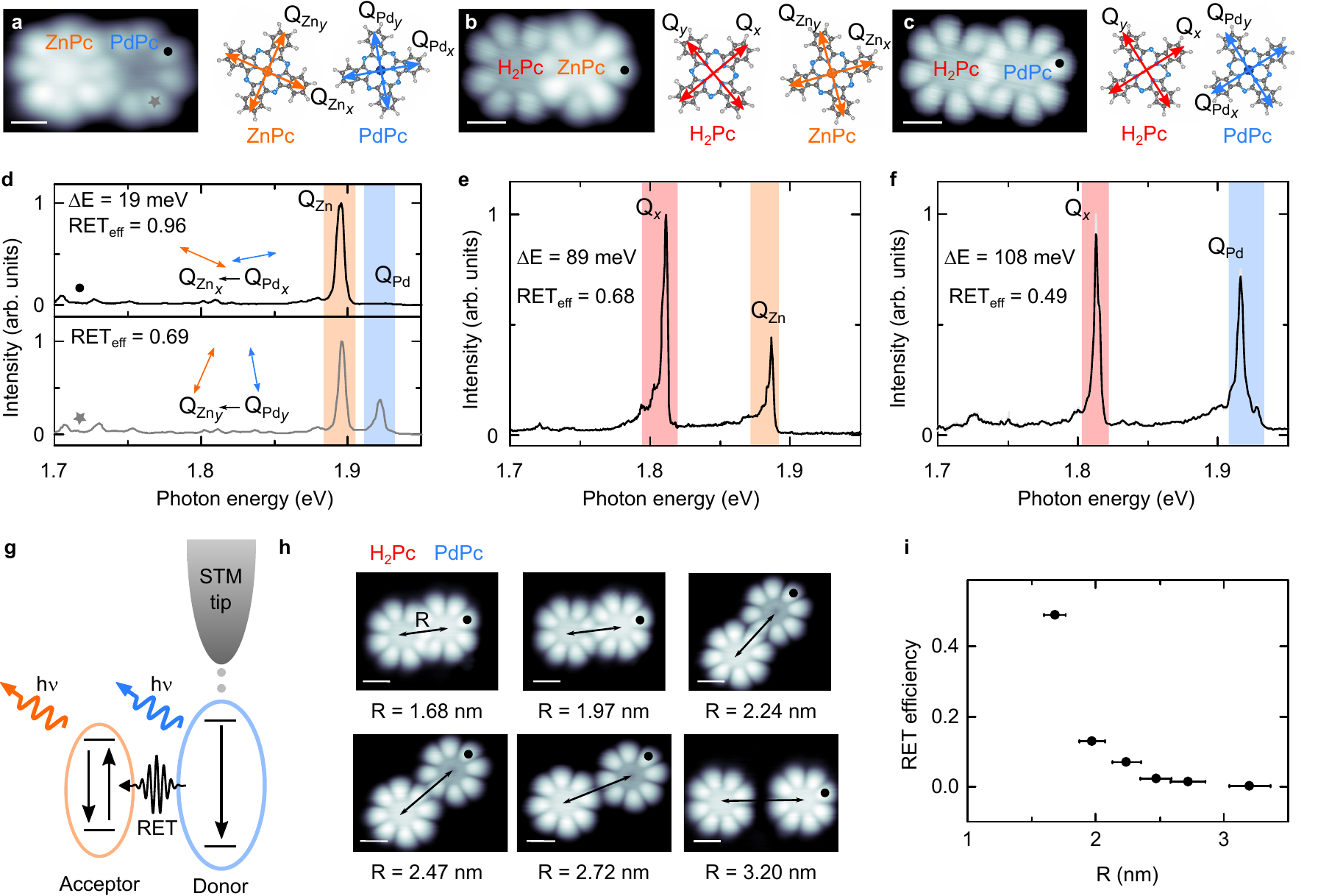}
  \caption{\label{fig2}  \textbf{RET between D--A pairs.} STM images ($V$ = -2.5 V) and models of the (a) PdPc--ZnPc ($I$ = 10 pA), (b) ZnPc--H$_{2}$Pc ($I$ = 10 pA) and (c) PdPc--H$_{2}$Pc ($I$ = 5 pA)  D--A pairs and (d, e, f) plasmon--corrected STML spectra ($V=$ -2.5 V). The markers in (a), (b) and (c) show the STM tip positions used to record the spectra in (d) ($I$ = 300 pA, acquisition time $t$ = 300 s), (e) ($I$ = 300 pA, $t$ = 180 s) and (f) ($I$ = 100 pA, $t$ = 180 s) respectively. The colored areas indicate the spectral range considered to calculate the $RET_{eff}$ values. 
  (g) Sketch of the D-A energy transfer experiments. (h) STM images ($I$ = 5 pA, $V$ = -2.5 V) of PdPc--H$_{2}$Pc dimers whose center-to-center distance (R) is progressively increased by STM-tip manipulation (two sets of data were used to generate this panel). (i) PdPc--H$_{2}$Pc RET efficiency as a function of $R$ deduced from STML spectra (\ExFigref{figure_rawdistance}) acquired for the STM-tip positions marked by dots in (h). Error bars are discussed in \ExFigref{figure_rawdistance}. Scale bars in STM images are 1 nm.} 
\end{figure} 

\begin{figure}
  \includegraphics[width=89 mm]{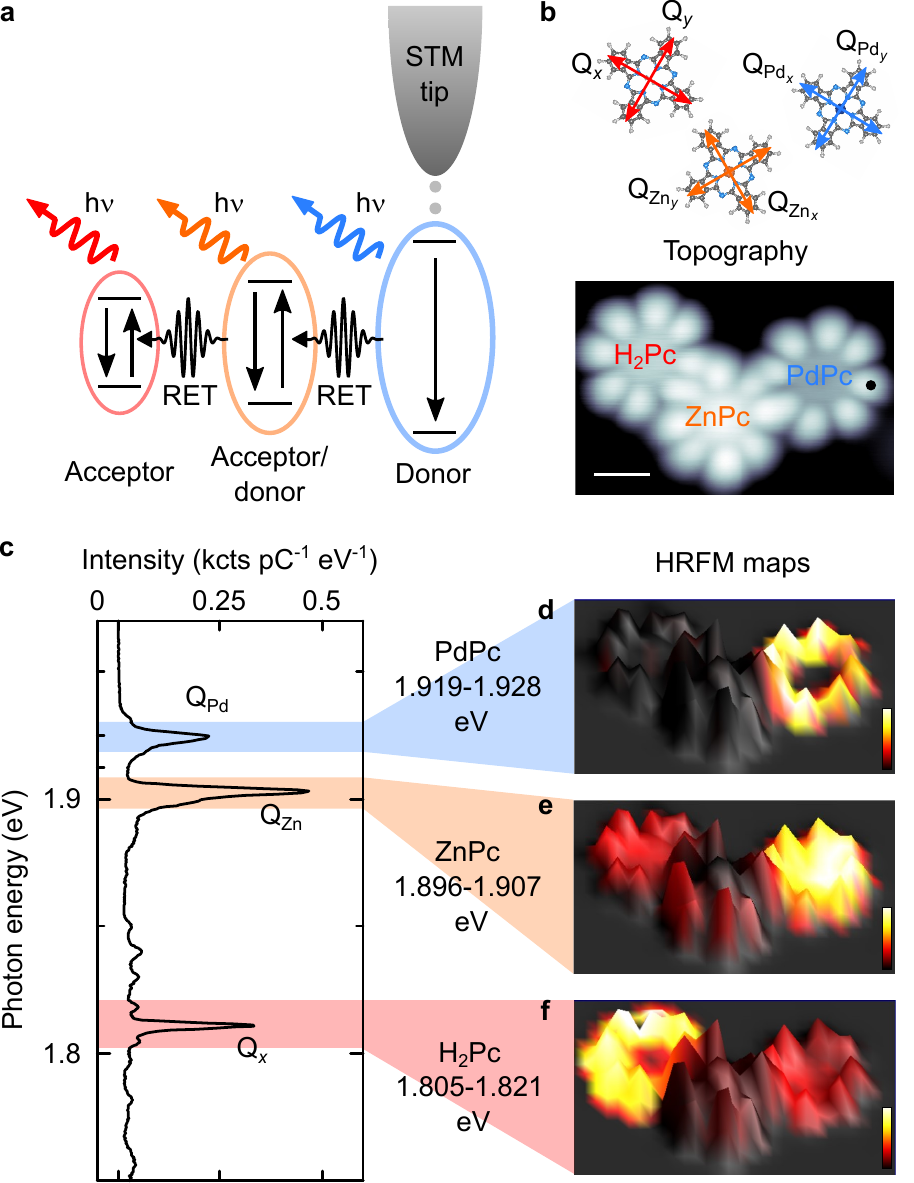}
  \caption{\label{fig3} \textbf{Cascaded RET} (a) Sketch of the experiment. (b) STM image ($I$ = 10 pA , $V$ = -2.5 V, scale bar 1 nm) and model of a PdPc--ZnPc--H$_{2}$Pc trimer. The black dot in the STM image indicates the STM-tip position used to acquire the STML spectrum ($I$ = 300 pA, $V=$ -2.5 V, $t$ = 200 s) in (c). HRFM maps ($I$ = 300 pA, $V=$ -2.5 V, time per pixel = 20 s) of the emission lines of (d) PdPc, (e) ZnPc and (f) H$_{2}$Pc. These maps reflect the emission intensity of a given chromophore as a function of the tip position. Emission intensity is coded in false colors and is overlaid on the pseudo-3D tunneling current image acquired simultaneously. The color scale range from (d) 0 to 0.1, (e) 0 to 0.3  and (f) 0 to 0.6 kcts pC$^{-1}$ eV$^{-1}$.} 
\end{figure}

In pioneering works, Zhang \textit{et al.} \cite{Zhang2016}, and Imada \textit{et al.} \cite{Imada2016} reported the emission of coherently coupled chromophores, and RET in a single donor--acceptor pair with STML.
Building on their approach, we first manipulated the molecules on NaCl to form the three possible donor-acceptor (D-A) pairs: PdPc--ZnPc (\Figref{fig2}a), ZnPc--H$_{2}$Pc (\Figref{fig2}b) and PdPc--H$_{2}$Pc (\Figref{fig2}c). In \Figref{fig2}d, e, and f we display for each pair the STML spectrum obtained by locating the STM tip on the extremity of the donor that is the most distant from the acceptor (black dots in the STM images of \Figref{fig2}). By this approach, the chances to excite directly the acceptor are negligible (see also the discussion of \Figref{fig3}d). These spectra systematically display peaks at both donor and acceptor energies, attesting that part of the energy of the donor is transmitted to the acceptor, a phenomenon described in terms of a RET process involving dipole-dipole interactions \cite{Imada2016}. Assuming perfect emitters, the energy transfer efficiency, $RET_{eff}$, (given in inset) follows: $RET_{eff} = I_{A}/(I_{D}+I_{A})$, where $I_{D}$ ($I_{A}$) is the emission intensity of the donor (acceptor) in the dimer. Qualitatively, the energy transfer is more efficient when the gap-energy difference between the donor and the acceptor is small. This behavior reflects the impact of a higher spectral overlap between the donor emission band and the acceptor absorption band. 
A semi-quantitative estimation of the spectral overlap for each D-A pair yields values of 6.3 (PdPc--ZnPc), 1.9 (ZnPc--H$_{2}$Pc) and 1.4 eV$^{-1}$ (PdPc--H$_{2}$Pc) (see Methods and \ExFigref{figS1}), following the trend observed in \Figref{fig2}. To ensure a fair comparison between the RET efficiency in the D-A pairs, the spectra of \Figref{fig2} were normalized by the plasmonic response of the tip-sample junction so as to correct for the different emission-line enhancements due to the more or less resonant character of the plasmon modes, and for the wavelength-dependent response of our detection setup\cite{Schull2009}.

RET relying on dipole-dipole interactions are also impacted by the angle between donor and acceptor dipoles \cite{Hinze2008}. To test this parameter, we show in \Figref{fig2}d spectra acquired for two STM tip positions (black dot and grey star in \Figref{fig2}a) on the donor of a PdPc--ZnPc dimer. These spectra reveal an efficient RET when the tip is located on top of the PdPc dipole oriented towards the ZnPc acceptor, a configuration where the donor Q$_{\textrm{Pd}_{x}}$ and acceptor Q$_{\textrm{Zn}_{x}}$ dipoles are nearly colinear (see Methods and \ExFigref{figureS2}). The RET is weaker when the dipoles are essentially parallel (Q$_{\textrm{Pd}_{y}}$ // Q$_{\textrm{Zn}_{y}}$). This implies that one can selectively excite a dipole by locating the tip on top of it, \textit{i.e.,} one can use an STM tip as an excitation source of sub-molecular-scale precision. A more detailed investigation (see Methods) gives a qualitative interpretation of the role played by the orientations of the donor and acceptor dipoles, but fails to provide quantitative predictions of the RET efficiency, suggesting that an accurate description of our RET processes cannot be resumed by simple Coulomb interactions between point dipoles.

To better establish this assertion, we evaluated (\Figref{fig2}h) the dependency of the RET process on the distance ($R$) between the centers of the donor and acceptor chromophores \cite{Andrews2011}. \Figref{fig2}i shows that the RET efficiency decreases monotonously when $R$ increases, and tends to zero for $R >$ 3.2 nm, approximately the distance at which the overlap of the donor and acceptor electronic orbitals vanishes in STM images. 
This supports that the RET mechanism goes beyond simple dipole-dipole interactions (\textit{i.e.}, F\"orster RET), and that other mechanisms occur at short D--A distances that rely on molecular orbitals overlap (\textit{i.e.}, exchange interactions or Dexter energy transfer) and/or multipolar RET\cite{Andrews2011}. 
Determining the respective influence of these different parameters requires detailed calculations that are beyond the scope of the present report.

Having characterized RET between two chromophores, we now focus on the decisive role played by intermediate chromophores to promote RET in multi-chromophoric systems. We first built a chromophores trimer (\Figref{fig3}a,b) where a large gap molecule (PdPc) is separated from a small gap molecule (H$_2$Pc) by an intermediate gap molecule (ZnPc). In this configuration, the central chromophore (ZnPc) may act as an ancillary molecule, accepting energy from PdPc before transferring it to H$_2$Pc. The fluorescence spectrum in \Figref{fig3}c, obtained for the tip located on PdPc (black dot in \Figref{fig3}b), confirms this picture. It reveals emission lines characteristic of the three chromophores, including the distant H$_2$Pc acceptor. The fact that an emission of light is also observed from the PdPc and ZnPc chromophores strongly hints towards sequential RET processes, where each chromophore is driven in its excited state before decaying either radiatively or by transferring its energy to a neighboring acceptor.

To further establish this mechanism, we realized HRFM maps \cite{Doppagne2020} (see Methods) for the three fluorescence lines (\Figref{fig3}d,e,f). They show the fluorescence intensity of a given molecule as a function of the excitation position, therefore providing a real space image of the sequential RET processes with atomic-scale resolution. \Figref{fig3}d shows that PdPc emits only when it is directly excited by the STM tip. This is expected considering that PdPc has the highest energy gap, but also confirms that a ''direct'' excitation of a molecule is impossible when the tip is located on a neighboring one. Interestingly, very few light is emitted when the tip excites the central part of PdPc, indicating a reduced coupling between tip plasmons and the molecular exciton compared to cases where the tip excites the periphery of the molecule \cite{Zhang2016,Kroger2018,Doppagne2020}.

\Figref{fig3}e shows an intense emission of ZnPc when the tip is located on PdPc, confirming the efficient RET observed in \Figref{fig2}d. The ZnPc emission is intense even when the tip is located on the center of PdPc. As this tip position does not favor an emission of PdPc (\Figref{fig3}d), almost all the energy stored in PdPc is transferred to ZnPc in this case. ZnPc emission is also observed for the tip located on H$_2$Pc, an effect similar to the one reported in [\onlinecite{Imada2016}], and associated to a transfer from the H$_2$Pc Q$_{y}$ to the lower energy Q$_\textrm{Zn}$. Surprisingly, the emission of ZnPc is weaker when the tip is directly located on top of it. Under this condition, a detailed analysis of the STML spectra of ZnPc (see Methods and \ExFigref{figcharge}) reveals that the tunneling current intermittently drives the molecule in its cationic form\cite{Doppagne2018}, ZnPc$^{+}$, leading to a reduced emission of the neutral molecule (h$\nu$ $\approx$ 1.90 eV). 
 
Eventually, \Figref{fig3}f reveals an intense emission of H$_2$Pc for a direct tip excitation, together with a rather large signal for the tip located on top of the most distant PdPc molecule (see also Methods and \ExFigref{figureS5}). This last observation reflects the sequential energy funneling from the large to the small gap molecule. Interestingly, one can compare the PdPc$\rightarrow$H$_2$Pc RET efficiency with and without the presence of the ancillary ZnPc molecule.
Despite the larger distance between the donor (PdPc) and the acceptor (H$_2$Pc), the imperfect alignment of the dipoles in the trimer and the energy lost by the emission of the ancillary molecule (ZnPc), the RET efficiency in the trimer (up to 0.33) is only slightly reduced compared to the dimer (0.49) for equivalent tip positions (black dots in STM images in \Figref{fig2}c and 3b). This demonstrates that sequential energy transfer can be used to funnel energy between distant centers with high efficiency, and shows that sequential RET processes with small energy steps may be more advantageous than single RET events with large energy gaps. This general result illustrates, at the level of three chromophores, the efficient strategy used by photosynthetic systems to convey energy from collection centers to distant reaction centers\cite{Mirkovic2017}, a concept also at play in artificial multi-chromophoric components \cite{hippius2008}.

\begin{figure}
  \includegraphics[width=89 mm]{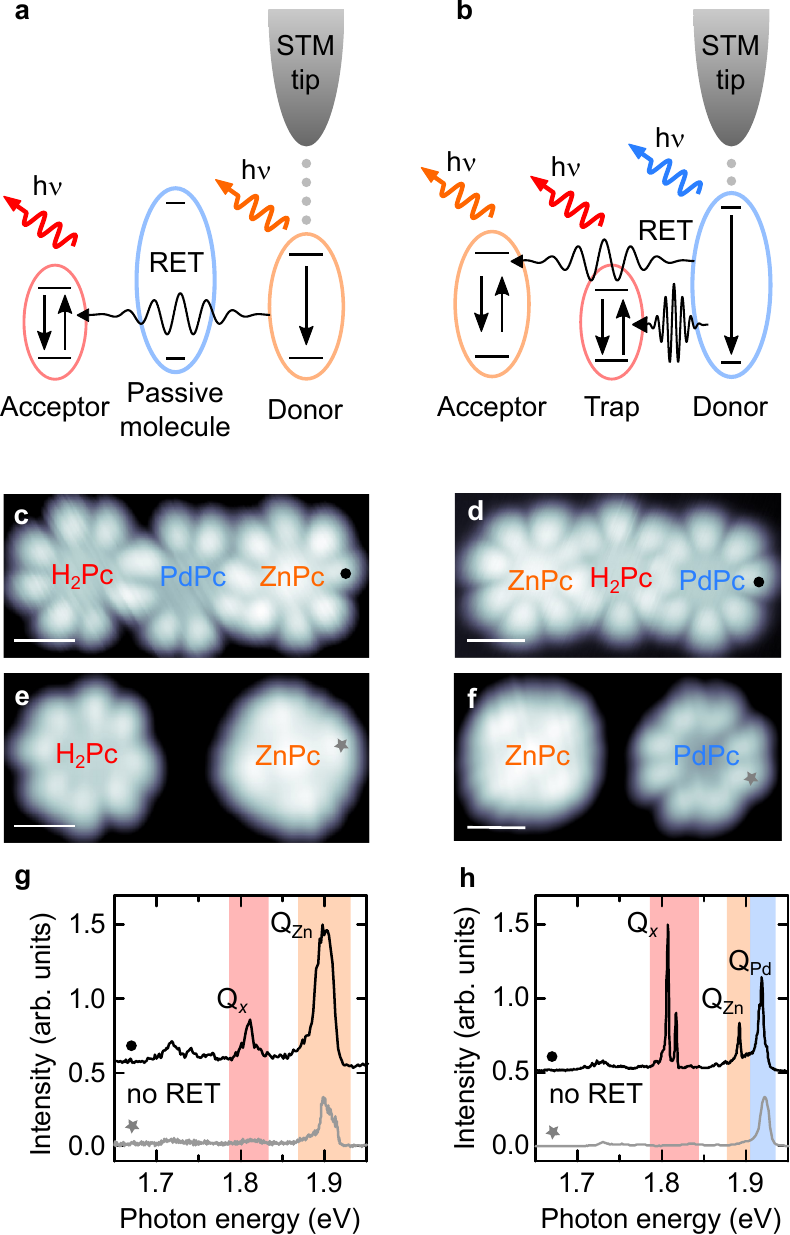}
 \caption{\label{fig4} \textbf{Promoting RET with passive and trap molecules} (a,b) Sketches of the experiments. (c-f) STM images ($V$ = - 2.5 V) of a ZnPc--PdPc--H$_{2}$Pc trimer (c, $I$ = 10 pA), a ZnPc--H$_{2}$Pc dimer separated by 3.7 nm (e, $I$ = 10 pA), a ZnPc--H$_{2}$Pc--PdPc trimer (d, $I$ = 5 pA) and a ZnPc--PdPc dimer separated by 3.1 nm (f, $I$ = 10 pA). Scale bars 1 nm. (g,h) STML spectra ($V$ = - 2.5 V) acquired at positions marked in (c)-(f). $I$ = 300 pA, $t$ = 120 s ($t$ = 300 s) for the black dot (grey star) in (g). $I$ = 200 pA, $t$ = 360 s ($I$ = 300 pA, $t$ = 100 s) for the black dot (grey star) in (h).} 
\end{figure} 

In other multi-chromophoric configurations (\Figref{fig4}), we have separated donors and acceptors by molecules having either a larger energy-gap than the donor (\Figref{fig4}a,c), which should therefore behave as passive elements, or a smaller energy-gap than the acceptor (\Figref{fig4}b,d), a configuration where they may act as excitation traps. For sake of comparison, we have used the same donor-acceptor pairs separated by vacuum over the same distance (\Figref{fig4}e,f). The first trimer shows that, despite its too large optical gap, the ''passive'' PdPc molecule promotes the ZnPc $\rightarrow$ H$_2$Pc RET that is absent when the passive molecule is replaced by vacuum (\Figref{fig4}e) . This configuration is very similar to RET experiments where a medium separates donors and acceptors, and highlights the role played by unexpected species that transiently pass between donors and acceptors in liquid experiments. Eventually, this process has its importance in RET occurring in synthetic D-A compounds separated by an organic bridge \cite{Hinze2008,curutchet2008} or in the antenna complex (LH2) of some photosynthetic bacteria, where passive carotenoid molecules promote energy transfer between donor and acceptor pigments \cite{Pullerits1997,Scholes2000}. This phenomenon has been investigated experimentally on ensembles of molecules and theoretically for individual elements similar to the one proposed here \cite{Andrews2013}. Whereas in a macroscopic picture it is the refraction index of the medium that modifies the efficiency of the RET process, it is the ac-polarizability of the passive molecule, allowing for a partial delocalization of the transition dipole moments of the donor and acceptor, that promotes dipole-dipole RET in our three-molecule model system. As the ac-polarizabitlity of a molecule primarily scales with its optical gap, we can conclude that PdPc, whose optical gap is very close to the one of the donor molecule (ZnPc in this case), is a very good choice to promote energy to the acceptor (see Methods and \ExFigref{figSI_model} for a coupled oscillator model). Besides, the passive molecule may also lead to a higher spatial overlap between donor and acceptor electronic orbitals by reducing the electronic attenuation factor, therefore increasing the efficiency of RET by superexchange coupling as it was reported for D-A pairs bridged by covalently linked organic units \cite{Pettersson2006}.

As a last experiment (\Figref{fig4}b), we tested the ability of an intermediate molecule of low energy-gap (H$_2$Pc) to act as a trap \cite{Mirkovic2017} preventing the diffusion of an excitation between distant donor (PdPc) and acceptor units (ZnPc) of higher energies. Here, the data reveal a strong emission of the intermediate (H$_2$Pc) molecule together with a weaker luminescence from the more distant ZnPc acceptor (\Figref{fig4}h). Hence, although most of the energy injected in the donor is trapped by H$_2$Pc, this molecule also promotes a direct RET channel from PdPc to ZnPc.  
Note (i) that RET mechanisms based on sequential charge transfers can be ruled out based on d$I$/d$V$ and ''at-distance'' measurements (see Methods, \ExFigref{figureS_didv} and \ExFigref{figureS_atdistance}), and (ii) that the Q$_y$ dipole of H$_2$Pc should play no role in the energy transfer pathway in \Figref{fig4}b,d,f,h, as its energy-gap lies higher than the one of PdPc.

Recently, several valuable approaches have been developed aiming at mimicking RET occurring in natural LHC with simplified systems such as microwave antennas\cite{rustomji2019}, quantum dots \cite{Kodaimati2018} or cold Rydberg atoms\cite{Ravets2014} assemblies. 
These approaches, however, fail to capture key properties intrinsic to organic systems, such as the influence of vibronic levels on (de-)coherence or the impact of molecular orbitals at close donor--acceptor distances. The original and most valuable aspect of our work is that it provides an extremely controlled approach to mimic energy funneling occurring in LHC by using the same elementary units, \textit{i.e.}, molecular chromophores. This allowed us to establish, at the single molecule limit, the key role played by ancillary and ''passive'' chromophores located between donors and acceptors to promote RET and funnel energy. Overall our molecular-scale system constitutes a unique platform to reproduce and probe RET mechanisms occurring in multichromophoric LHC with ultimate chemical, spatial and spectral precision. Future studies will investigate the dynamics of the RET process in STM junctions, establish the possible role of coherence between interacting chromophores, determine the influence of the plasmons localised at the tip-sample junction and decipher the influences of dipolar, multipolar and energy-exchange interactions on the RET mechanism.         

\noindent

\bibliographystyle{naturemag}

\section*{Methods}
              
\subsection*{Experimental}
The STM data were acquired with a low temperature (4.5\,K) Omicron setup operating in ultrahigh vacuum adapted to detect the light emitted at the tip-sample junction. The optical detection setup\cite{Chong2016} is composed of a spectrograph coupled to a CCD camera and provides a spectral resolution of $\approx$ 1 nm. Tungsten STM-tips were introduced in the sample to cover them with silver so as to tune their plasmonic response. The Ag(111) substrates were cleaned with successive sputtering and annealing cycles. Approximately 0.5 monolayer of NaCl was sublimed on Ag(111) kept at room temperature. The sample is then flash-annealed up to 370 K to obtain square domains of bi- and tri-layers of NaCl. Eventually, ZnPc, PdPc and H$_{2}$Pc molecules were evaporated on the cold ($\approx$ 5 K) NaCl/Ag(111) sample in the STM chamber. ZnPc and H$_{2}$Pc molecules were purchased from Sigma--Aldrich, while PdPc molecules were synthesized according to the procedure described in [\onlinecite{Lokesh2013}].

The ZnPc molecules were manipulated by positioning the STM tip at the edge of the molecule at a bias $V$ = +2.5 V or $V$ = -2.5 V and then reducing the tip-molecule distance until a molecular jump occurs\cite{Zhang2016}. The H$_{2}$Pc and PdPc were manipulated using voltage pulses (of up to -4 V at a set point of $V$ = -2.5 V and $I$ = 10 pA) applied at the edge of the molecule.

HRFM maps presented in \Figref{fig3} were generated by scanning the molecular trimer with the STM tip while recording a STML spectrum for each tip position (that is, pixel of the map). The tip--substrate separation was kept constant during the acquisition (open feedback loop) to prevent any distance-related artefacts. To compensate for the tip drift during the long acquisition of the HRFM maps, the ($x$,$y$,$z$) position of the tip was corrected using an ''atom-tracking'' procedure between the acquisition of each pixel, similarly to the procedure developed in  [\onlinecite{Kawai2011}]. The next step consisted in choosing the photon energy windows corresponding to the spectral features of interest (\textit{e.g.}, Q$_\textrm{Pd}$, Q$_\textrm{Zn}$ and Q$_{x}$ in \Figref{fig3}) to generate the corresponding photon intensity maps. All HRFM maps were recorded simultaneously in a single experimental run and could be readily compared. For each pixel of the maps, the photon intensity was normalized by the STM current recorded simultaneously so as to consider an equivalent excitation source for each pixel. \\ 

\subsection*{Estimation of the spectral overlaps for the donor--acceptor dimers}

\ExFigref{figS1} shows the fluorescence (F$_{D}$) and absorption ($\epsilon_{A}$) spectra for the donors (D) and acceptors (A) forming dimers in Fig.\,2 of the main manuscript. The donor fluorescence spectra correspond to the single molecule STML spectra presented in Fig.\,1 of the main manuscript. The acceptor absorption spectra are obtained by mirroring the fluorescence spectra at the maximum intensity energy, assuming a negligible Stokes shift as is generally expected for low temperature spectra of rigid phthalocyanine molecules  \cite{Murray2011}. Note that for the H$_{2}$Pc molecule we neglect the intensity of the Q$_{y}$ mode as its energy is higher than Q$_\mathrm{{Pd}}$ and Q$_\mathrm{{Zn}}$. The spectra are normalized such that:

\begin{equation}
      \label{eq:emission}
      \int_{0}^{\infty} F_{D}(E) dE = 1
\end{equation}
\begin{equation}
      \label{eq:absorption}
      \int_{0}^{\infty} \epsilon_{A}(E) dE = 1
\end{equation}

The normalized spectra are then used to calculate the spectral overlap $J$ as defined for the exchange interaction \cite{Speiser1996}:

\begin{equation}
      \label{eq:overlap}
      J = \int_{0}^{\infty} F_{D}(E)\epsilon_{A}(E) dE
\end{equation}

The $J$ values for the three studied D--A pairs, together with the energy differences between D and A main emission lines ($\Delta E$), and the RET efficiency values are reported in \ExFigref{figS1}d. Similarly to what is observed with other methods, this table shows that the RET efficiency in STM junctions primarily depends on the spectral overlap between the donor and the acceptor. This table also shows that the energy difference ($\Delta E$) between the donor and acceptor emission lines can be used as an acceptable and convenient approximation to intuit the energy transfer efficiency at low temperature between a donor and an acceptor. However, as the RET efficiency is also affected by several other parameters (D--A distance, dipole--dipole orientations...) that are slightly different from one D--A pair to the other, one does not find the expected proportional dependency of the RET efficiency with $J$.

\subsection*{The effect of the relative dipole orientation on RET}

In the F\"orster description of RET, the relative orientation of the donor and acceptor dipoles plays a crucial role on the efficiency of the energy transfer process.  It is reflected in the dipole orientation factor $\kappa^{2}$ defined as \cite{Wong2004}:
\begin{equation}
      \label{eq:kappa2}
      \kappa^{2} = (\sin \theta_{D} \sin \theta_{A} \cos \phi - 2 \cos \theta_{D} \cos \theta_{A})^2
\end{equation}
with the angles $\theta_{D}$ and $\theta_{A}$ as defined in \ExFigref{figureS2}a. $\phi$ is the dihedral angle between the planes of the dipoles, which in our case (dipoles in the same plane) is $\phi =0$.
Formula \ref{eq:kappa2} is derived under the assumption of point dipoles.

We calculated $\kappa^{2}$ (\ExFigref{figureS2}b) for the PdPc--ZnPc dimer presented in Fig. 2a of the main manuscript. Here, it is assumed that placing the tip at the extremity of one of the pyrrole sub-unit of the donor results in the specific excitation of the corresponding dipole (Q$_{\textrm{Pd}_{x}}$ or Q$_{\textrm{Pd}_{y}}$). From there, the transfer of energy to both dipoles of the acceptor (Q$_{\textrm{Zn}_{x}}$ and Q$_{\textrm{Zn}_{y}}$) is considered.

We first analyze the excitation of Q$_{\textrm{Pd}_{x}}$  (black dot on \ExFigref{figureS2}c), which results in a very efficient energy transfer (RET$_{eff}$ = 0.96) to ZnPc (black spectrum in \ExFigref{figureS2}d and \ExFigref{figureS2}b). Here, the Q$_{\textrm{Pd}_x}$ and Q$_{\textrm{Zn}_x}$ dipoles are nearly aligned, corresponding to an optimum RET configuration as confirmed by a $\kappa^{2}$ value close to $\kappa^{2}_{max}$ = 4. The estimation of $\kappa^{2}$ for the Q$_{\textrm{Pd}_x}$ $\xrightarrow{}$ Q$_{\textrm{Zn}_{y}}$ energy transfer leads to a much smaller value (0.43) indicating a negligible contribution of this RET path.

Excitation of Q$_{\textrm{Pd}_{y}}$ (gray star in \ExFigref{figureS2}c) results in a less efficient RET process (gray spectrum in \ExFigref{figureS2}e and \ExFigref{figureS2}b). Here, the RET process is strongly dominated by the close-to-parallel configuration of the Q$_{\textrm{Pd}_{y}}$ $\xrightarrow{}$ Q$_{\textrm{Zn}_{y}}$ RET path, leading to a $\kappa^{2}$ value of 1.06.

Overall, the PdPc--ZnPc dimer allows one to compare the dependence of the RET efficiency with dipole-dipole orientations for close-to ideal inline and parallel configurations. In agreement with the F\"orster theory, the former leads to a more efficient energy transfer than the latter, though only by a factor $\approx$ 1.4, while a factor of $\approx$ 4 is expected from the estimation of $\kappa^{2}$ in equation 4. The very short distance separating the donor and the acceptor in our experiment explains this quantitative discrepancy. %

Indeed, the F\"orster approach of RET holds only for D--A distances
much larger than the sizes of the chromophores, which is not the case in our experiment. For such short D--A distances, the point dipole approximation does not hold \cite{Speiser1996}, the dipole orientations may
be affected by the proximity with molecular neighbors \cite{Wong2004}; multipolar contributions to the RET
can no longer be neglected, and exchange energy process (\textit{i.e.}, Dexter energy transfer) may play a significant role.

Considering the influence of all these parameters, it is remarkable that the variation of the RET efficiency with the tip position in our STML experiment can be qualitatively accounted for by considering the F\"orster approach. 
Additional discussion on the influence of the dipole alignment on the RET process can be found in the "RET efficiency maps" section and \ExFigref{figureS5}. 

\subsection*{Charge state of the ZnPc molecule}

\ExFigref{figcharge}b and c compare large energy range STML spectra obtained from a ZnPc molecule in a PdPc--ZnPc dimer for positions marked in the STM image in \ExFigref{figcharge}a. The spectrum in \ExFigref{figcharge}b corresponds to a direct excitation of the molecule (\textit{i.e.}, the tip is located on top of the ZnPc molecule), while the spectrum in \ExFigref{figcharge}c displays the indirect excitation of the ZnPc fluorescence (\textit{i.e.} the tip is located on top of the PdPc molecule and the luminescence of ZnPc results from the PdPc $\xrightarrow{}$ ZnPc energy transfer). As discussed in the main manuscript, an indirect excitation leads to a $13$ times more efficient excitation of the ZnPc fluorescence than a direct excitation (note the different vertical scales in \ExFigref{figcharge}b and c). 

The spectrum in \ExFigref{figcharge}b also reveals a spectral feature at $\approx$ 1.52\,eV that is absent in the spectrum of \ExFigref{figcharge}c. In a recent STML experiment performed with ZnPc molecules deposited on a NaCl/Au(111) surface, this spectroscopic contribution was assigned to the emission of ZnPc$^+$, the positively charged molecule \cite{Doppagne2018}. In this paper, it was concluded that the tunneling current traversing the molecule is responsible for a fast switching between the neutral and cationic states during STML spectra acquisition. This leads to the observation of both the charged and neutral contributions in the STML spectra of ZnPc/NaCl/Au(111). The spectrum in \ExFigref{figcharge}b indicates a similar behavior for ZnPc/NaCl/Ag(111), although with a lower relative intensity of the charged peak with respect to the neutral one compared to Au(111).

In contrast, the charged contribution is absent in the STML spectrum of \ExFigref{figcharge}c,  corresponding to an indirect excitation of the ZnPc, confirming that the tunneling current must pass through the ZnPc molecule to charge it. In turn, the molecule spends a smaller fraction of time in the neutral state for a direct excitation, leading thus to a reduced emission intensity at $h\nu \approx$  1.90\,eV, compared to the indirect excitation spectrum (\ExFigref{figcharge}b,c). It also explains why the ZnPc molecule appears dimmer than the H$_2$Pc and PdPc molecules in the HRFM map of Fig.\,3e. 

\subsection*{RET efficiency maps} 

Based on the HRFM maps presented in Fig.\,3d,\,e,\,f of the main manuscript, one can reconstruct RET efficiency maps by applying the formula $RET_{eff} = I_{A}/(I_{PdPc}+I_{ZnPc}+I_{H_{2}Pc})$ to each pixel of the map. The result of this operation is presented in \ExFigref{figureS5}a and b for H$_{2}$Pc and ZnPc molecules acting both as acceptors. 

By displaying normalized photon intensities, these maps reveal the RET signals from molecules having low excitation probabilities, such as ZnPc, which are too weak to be visualized in HRFM maps. For example in \ExFigref{figureS5}a, one observes a high RET efficiency to the H$_{2}$Pc molecule when the tip is located on the adjacent ZnPc donor, a behavior that is hidden in the HRFM map in Fig.\,3e.

Interestingly, the impact of the dipole alignment on the RET is here readily visible: the excitation of Q$_{\textrm{Zn}_{x}}$, which is well-aligned with Q$_{x}$, leads to a quite efficient ($RET_{eff} > 0.5$) energy transfer to the H$_{2}$Pc acceptor. In contrast, the excitation of Q$_{\textrm{Zn}_{y}}$ results in a low $RET_{eff}$ due to the near-parallel orientation of the Q$_{\textrm{Zn}_{y}}$ and Q$_{{y}}$ dipoles (see "The effect of the relative dipole orientation on RET" section and \ExFigref{figureS2} for more details on the impact of dipole--dipole angles on $RET_{eff}$).

A similar discussion can be proposed for the RET efficiency map from PdPc to ZnPc (\ExFigref{figureS5}b). Here, the RET efficiency is equivalent for the STM tip located on top of the two donor dipoles, Q$_{\textrm{Pd}_{x}}$ and Q$_{\textrm{Pd}_{y}}$. This is a consequence of the specific dipole alignment of the donor and acceptor molecules: the Q$_{\textrm{Pd}_{x}}$ and Q$_{\textrm{Zn}_{y}}$ dipoles, and the Q$_{\textrm{Pd}_{y}}$ and Q$_{\textrm{Zn}_{y}}$ dipoles make an angle of $\approx 45^\circ$.

In conclusion, RET efficiency maps provide interesting complementary information to HRFM maps by highlighting the role of dipole orientations and allowing to visualize RET from molecular donors having a low excitation probability.

\subsection*{Excluding charge transfer mechanisms} 

In their recent publication on energy transfer \cite{Imada2016} Imada \textit{et al.} compared d$I$/d$V$ data acquired on both isolated and paired donor/acceptor molecules to exclude an energy transfer mechanism based on sequential charge transfer (CT) steps. In \ExFigref{figureS_didv}, we produce similar data acquired for individual chromophores and for the trimer of \Figref{fig4}c. The orbital positions in the d$I$/d$V$ spectra of individual H$_{2}$Pc, PdPc, and ZnPc (\ExFigref{figureS_didv}a) are very similar to the ones deduced from d$I$/d$V$ spectra measured along a line crossing a H$_{2}$Pc--PdPc--ZnPc trimer (\ExFigref{figureS_didv}b), indicating that there is no measurable hybridization between the ground state orbitals of the chromophores. Following the argument developed in [\onlinecite{Imada2016}], we note that the highest occupied orbital of the ZnPc donor is at a higher energy than the one of the PdPc passive and H$_{2}$Pc acceptor molecules, preventing a migration of holes that would be injected with the STM tip in the ZnPc molecule, and excluding a CT mediated energy transfer mechanism. Another, more straightforward, evidence that  CT mechanisms are not responsible for the RET processes is provided in \ExFigref{figureS_atdistance}, where we demonstrate RET in the absence of charge injection into the donor. For that, we located the tip at a $\approx$ 2.2 nm distance from the center of the PdPc molecule of a PdPc-H$_{2}$Pc donor-acceptor pair  and generate a so-called ''at-distance'' excitation of the donor \cite{Imada2017,Zhang2017,Kroger2018}. Once normalized by a plasmon reference spectrum ($I_{0}$), this STML spectrum (\ExFigref{figureS_atdistance}c) reveals the signature of both PdPc and H$_{2}$Pc molecules. As no charge is running through the pair in this case, a contribution of CT to the RET mechanism can be safely ruled out.

\subsection*{Modeling  the  role  of  an intermediate  molecule  with  a  classical  oscillatory approach}

The energy transfer between a donor (D) and an acceptor (A) molecule via an intermediate molecule (I) (\Figref{fig3} and \Figref{fig4}) is illustrated by a coupled harmonic oscillator model (\ExFigref{figSI_model}) composed of three pendulums, where the donor pendulum (D) is coupled to an acceptor pendulum (A) through an intermediate pendulum (I). 
The coupled oscillator equations describing the time evolution of the system read

\begin{eqnarray}
  \ddot{\theta}_D\left(t\right)&=& -\left(\omega_D^2 +\omega_1^2\right)\theta_D\left(t\right) + \omega_1^2\theta_I\left(t\right) - \frac{\omega_D}{Q_D}\dot{\theta}_D\left(t\right)  \nonumber \\
 \ddot{\theta}_I\left(t\right)&=& -\left(\omega_I^2+\Omega_1^2+\Omega_2^2\right)\theta_I\left(t\right)+\Omega_1^2\theta_D\left(t\right)+\Omega_2^2\theta_A\left(t\right)- \frac{\omega_I}{Q_I}\dot{\theta}_I\left(t\right),\nonumber\\ 
  \ddot{\theta}_A\left(t\right)&=& -(\omega_A^2 +\omega_2^2)\theta_A\left(t\right) + \omega_2^2\theta_I\left(t\right) - \frac{\omega_A}{Q_A}\dot{\theta}_A\left(t\right),  \nonumber\\
  \Omega_1^2&=&\frac{1}{\alpha_D}\frac{\omega_1^2\omega_I^4}{\omega_D^4},\nonumber\\
  \Omega_2^2&=&\frac{1}{\alpha_A}\frac{\omega_2^2\omega_I^4}{\omega_A^4}.\nonumber
\end{eqnarray}

\noindent
where the $\theta_i\left(t\right), i=D,I,A$ are the angles of the donor, intermediate and acceptor pendulums  with respect to their equilibrium positions at time $t$.
$\alpha_D=m_I/m_D$ and $\alpha_A=m_I/m_A$ are the mass ratios;  in this paper the masses are considered identical for all pendulums so that $\alpha_D=\alpha_A=1$.
$\omega_D$,  $\omega_I$,  $\omega_A$ are the eigenfrequencies of the uncoupled pendulums.
Coupling constants between (D) and (I), and (I) and (A) are $\omega_1$,  $\omega_2$, respectively; note that, in this model, there is no direct coupling between pendulum (D) and (A), so that pendulum (I) acts as a relay of the excitation.
We consider damping with quality factors  $Q_D$, $Q_I$, $Q_A$, taken all identical for the simulations ($Q=60$).

All pendulums are in their equilibrium positions ($\theta = 0$ rad) at $t = 0$.
To mimic the excitation of the donor by the tunneling current, pendulum (D) is provided with an initial angular speed ($\dot{\theta}_D(t=0)=1\ \mathrm{rad}/s$); the initial angular speed of (I) and (A) are zero.

The time evolution of the system is computed using a velocity-Verlet algorithm\cite{verlet1967}. The angles at time $t+\delta t$ are computed from their values at time $t$ by using

\small{
\begin{eqnarray}
  \theta_D\left(t+\delta t\right)&=& \left\{1-\frac{\delta t^2}{2}\left(\omega_D^2+\omega_1^2\right)   \right\}\theta_D\left(t\right)+\frac{\delta t^2}{2}\omega_1^2\theta_I\left(t\right)+ \left(1-\frac{\delta t}{2}\frac{\omega_D}{Q_D} \right)\delta t \dot{\theta}_D\left(t\right),\nonumber  \\
  \theta_I\left(t+\delta t\right)&=& \left\{1-\frac{\delta t^2}{2}\left(\omega_I^2+\Omega_1^2+\Omega_2^2\right)   \right\}\theta_I\left(t\right)+\frac{\delta t^2}{2}\Omega_1^2\theta_D\left(t\right)+\frac{\delta t^2}{2}\Omega_2^2\theta_A\left(t\right) +\left(1-\frac{\delta t}{2}\frac{\omega_I}{Q_I} \right)\delta t \dot{\theta}_I\left(t\right), \nonumber\\
  \theta_A\left(t+\delta t\right)&=& \left\{1-\frac{\delta t^2}{2}\left(\omega_A^2+\omega_2^2\right)   \right\}\theta_A\left(t\right)+\frac{\delta t^2}{2}\omega_2^2\theta_I\left(t\right)+ \left(1-\frac{\delta t}{2}\frac{\omega_A}{Q_A} \right)\delta t \dot{\theta}_D\left(t\right).\nonumber
\end{eqnarray}}

Then, the values of the angular speeds at time $t+\delta t$ are computed from their values at time $t$ and from the values of the angles at times $t$ and $t+\delta t$ by using

\small{
\begin{eqnarray}
  \dot{\theta}_D\left(t+\delta t\right)&=& \left(
  -\frac{\delta t}{2}\left(\omega_D^2+\omega_1^2\right)\left(\theta_D\left(t+\delta t\right)+\theta_D\left(t\right)\right)
                                           \right.\nonumber\\
                                           &&\left.+\frac{\delta t}{2}\omega_1^2\left(\theta_I\left(t+\delta t\right)+\theta_I\left(t\right)\right)
                                           + \left(1-\frac{\delta t}{2}\frac{\omega_D}{Q_D} \right) \dot{\theta}_D\left(t\right)\right)
                                           \left(1+\frac{\delta t}{2}\frac{\omega_D}{Q_D} \right)^{-1},\nonumber  \\
  \dot{\theta}_I\left(t+\delta t\right)&=&\left(
                                           -\frac{\delta t}{2}\left(\omega_I^2+\Omega_1^2+\Omega_2^2\right)\left(\theta_I\left(t+\delta t\right)+\theta_I\left(t\right)\right)
                                           +\frac{\delta t}{2}\Omega_1^2\left(\theta_D\left(t+\delta t\right)+\theta_D\left(t\right)\right)\right.\nonumber \\
                                           &&\left. +\frac{\delta t}{2}\Omega_2^2\left(\theta_A\left(t+\delta t \right) +\theta_A\left(t\right)\right)+\left(1-\frac{\delta t}{2}\frac{\omega_I}{Q_I} \right)\dot{\theta}_I\left(t\right)\right)\left(1+\frac{\delta t}{2}\frac{\omega_I}{Q_I} \right)^{-1}, \nonumber \\
  \dot{\theta}_A\left(t+\delta t\right)&=& \left(
  -\frac{\delta t}{2}\left(\omega_A^2+\omega_2^2\right)\left(\theta_A\left(t+\delta t\right)+\theta_A\left(t\right)\right)
                                           \right.\nonumber\\
                                           &&\left.+\frac{\delta t}{2}\omega_2^2\left(\theta_I\left(t+\delta t\right)+\theta_I\left(t\right)\right)
                                           + \left(1-\frac{\delta t}{2}\frac{\omega_A}{Q_A} \right) \dot{\theta}_A\left(t\right)\right)
                                           \left(1+\frac{\delta t}{2}\frac{\omega_A}{Q_A} \right)^{-1}.\nonumber  
\end{eqnarray}}

By repeating this procedure several thousand times, one calculates how the initial excitation of (D) is transferred to (I) and (A). In the following, we will show how this transfer of excitation varies as a function of the resonance frequency of (I). As such, this simple classical model can be used to infer the impact of the intermediate molecule energy-gap size on the RET efficiency in the molecular structures investigated in \Figref{fig3} and \Figref{fig4}.

In \ExFigref{figSI_model}b,c,d we represent the oscillation amplitudes of the three pendulums as a function of the normalized time $t/T_D$ (where $T_D = 2\pi/\omega_D$), for a large (b), a medium (c) and a low (d) eigenfrequency of the intermediate pendulum. $20000$ steps with a time step $\delta t=T_D/300$ were necessary to generate the time evolution of the system. 
We also estimate the fraction of the energy dissipated by each pendulum as a function of the normalized intermediate pendulum eigenfrequency $\omega_I/\omega_D$ by integrating over the time the work of the friction forces for  each oscillator (\ExFigref{figSI_model}e). In panels (b) and (e), we see that a very small fraction of the original excitation is dissipated in the intermediate pendulum (\textit{i.e.}, low amplitude of the blue oscillations), while a larger fraction is dissipated in the acceptor pendulum (\textit{i.e.}, larger amplitude of the red oscillations), reflecting the ability of a ''passive'' molecule to promote RET processes between a donor and an acceptor.
When the intermediate pendulum eigenfrequency is considered in-between the donor and acceptor ones (\ExFigref{figSI_model}d,e), the energy dissipated in the intermediate pendulum increases strongly, which is expected considering the near-resonant condition. The energy dissipated in the acceptor pendulum increases as well, confirming that the efficiency of the RET process is improved when the intermediate molecule can serve as an ancillary unit. Eventually, the case of a low-eigenfrequency intermediate pendulum is investigated (d), revealing a weak dissipation by the intermediate and acceptor pendulums, somehow supporting the idea that low-energy gap molecules can be used to block a RET path.
 
While this simple classical model provides an intuitive understanding of the role played by the energy-gap of the intermediate molecule, it fails to capture the quantum nature of the energy transfer process (\textit{e.g.}, exchange interactions, which reflects a Dexter process, are not accounted for by this approach). Such effects require a time-dependent quantum approach, which is beyond the scope of this study.

\section*{Data availability}
The data that support the plots within this paper and other findings of this study are available from the corresponding authors (A.R. and G.S.) upon reasonable request.

\section*{Acknowledgements}
The authors thank Virginie Speisser for technical support and Alex Boeglin for discussions. This project has received funding from the European Research Council (ERC) under the European Union's Horizon 2020 research and innovation program (grant agreement No 771850) and the European Union's Horizon 2020 research and innovation programme under the Marie Sklodowska-Curie grant agreement No 894434. The Agence National de la Recherche (project Organiso No. ANR-15-CE09-0017), the Labex NIE (Contract No. ANR-11-LABX-0058\_NIE), and the International Center for Frontier Research in Chemistry (FRC) are acknowledged for financial support. 

\section*{Author contributions}
S.C., A.R., B.D. and G.S. conceived, designed and performed the experiments. S.C., A.R., M.R., F.S. and G.S. analysed the experimental data. F.S. and H.B. performed the oscillatory model. M.F. and F.C. synthesised the PdPc chromophores. All the authors discussed the results and contributed to the redaction of the paper. 

\newpage
\noindent

\renewcommand{\figurename}{Extended Data Fig.}
\renewcommand\thefigure{\arabic{figure}}    
\setcounter{figure}{0}

\begin{figure*}[!h]
\centering
\includegraphics[width=0.95\linewidth]{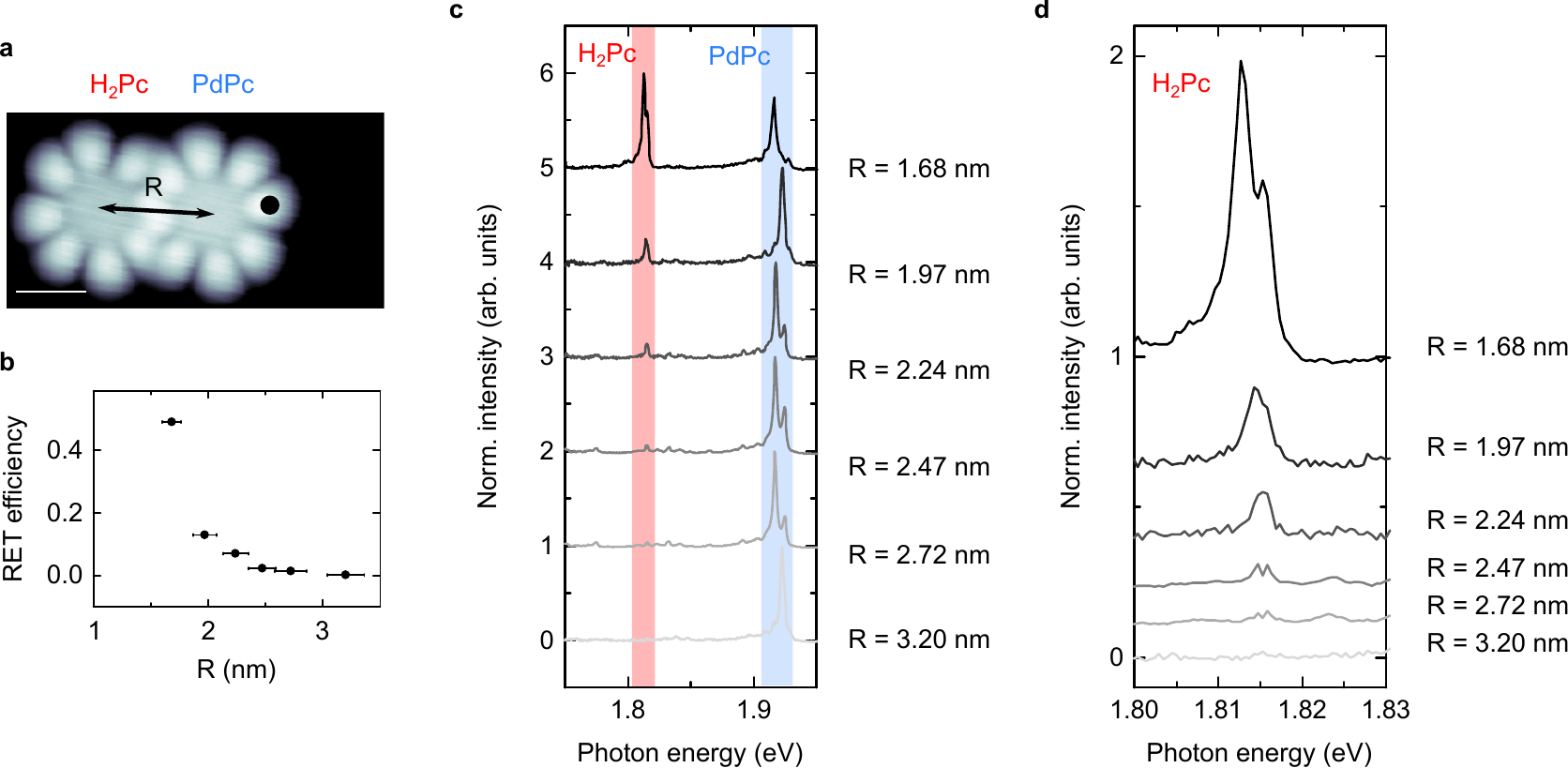}\caption{Spectra used to obtain the D--A distance dependence presented in Fig.\,2i of the main manuscript. (a) STM image of the H$_{2}$Pc--PdPc dimer, $I = 5$\,pA, $V = -2.5$\,V. Scale bar 1 nm. (b) $RET_{eff}$ values calculated from the spectra in (c). The horizontal error bars consider a 5 \% error in the estimation of the molecules positions. The vertical error bars are smaller than the symbol size as the statistical error on the number of photon counts is less than 3 \%.  (c) Normalized STML spectra acquired at positions marked in Fig.\,2h, $V = -2.5$\,V, acquisition time $t = 60$\,s, $I = 50$\,pA (for $R = 2.24$\,nm), $I = 100$\,pA (for $R = 1.68, 1.97, 3.2$\,nm), $I = 250$\,pA (for $R = 2.47, 2.72$\,nm). The spectra were normalized by the plasmonic response of the cavity \cite{Reecht2014} to ensure a fair comparison between the intensities of the molecular emission lines, and scaled to unity. The splits observed in the PdPc spectra correspond to a partial lifting of the Q$_{\mathrm{Pd}_{x}}$ and Q$_{\mathrm{Pd}_{y}}$ degeneracy that seems to depend on details of the adsorption site. A similar effect has been reported previously \cite{Doppagne2020} for H$_{2}$Pc. The $RET_{eff}$ values were estimated by integrating the light emission intensities in the spectral ranges indicated in blue and red. (d) Enlarged view of the H$_2$Pc line displayed in (c). }
\label{figure_rawdistance}
\end{figure*}

\newpage
\noindent

\begin{figure}[h]
\centering
  \includegraphics[width=0.9\linewidth]{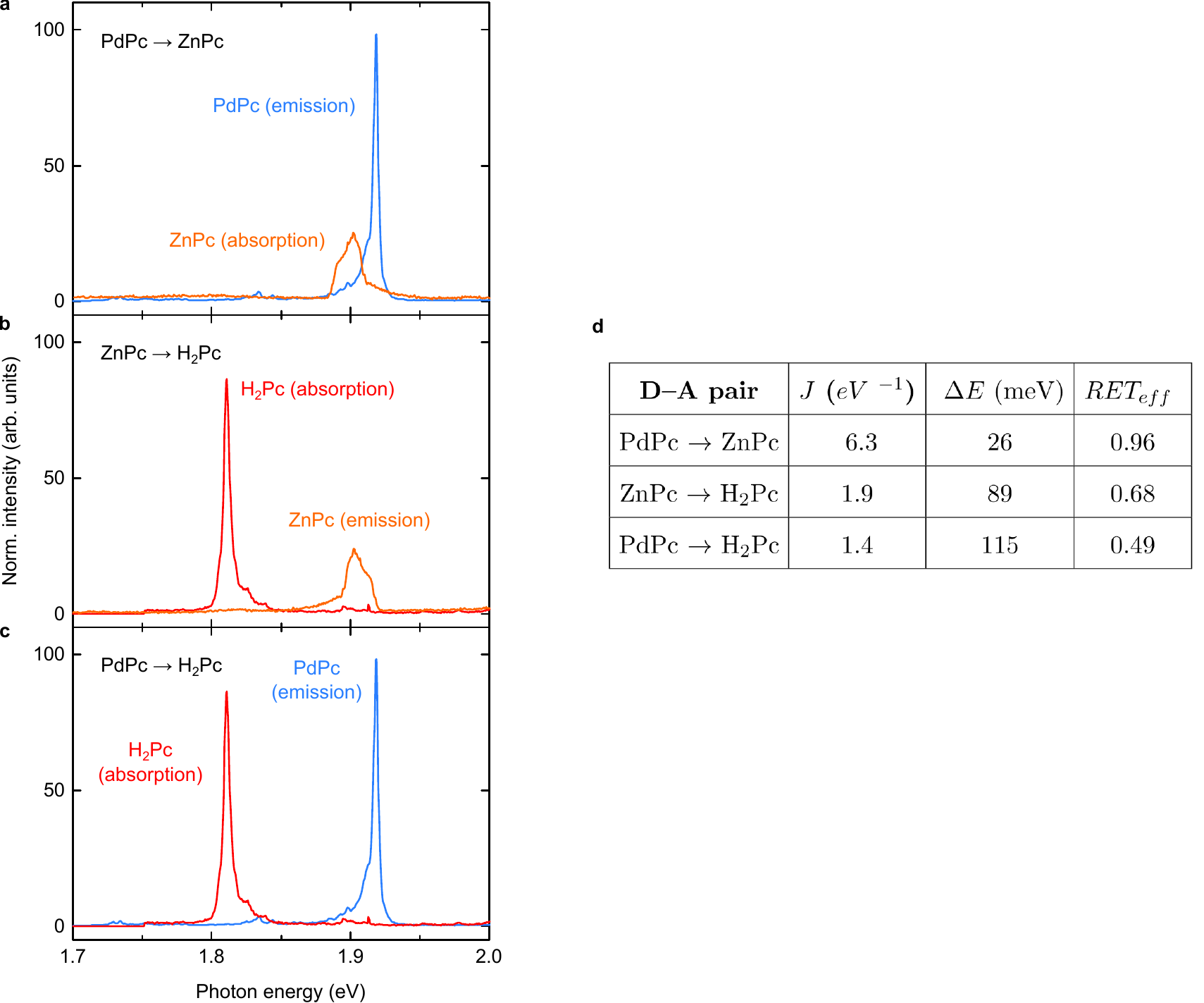}
  \caption{\label{figS1} Normalized emission and absorption spectra for (a) PdPc--ZnPc (b) ZnPc--H$_{2}$Pc and (c) PdPc--H$_{2}$Pc dimers. (d) Spectral overlaps ($J$), energy differences ($\Delta E$) and RET efficiencies for the dimer configurations presented in Fig.\,2.}      
\end{figure}

\newpage
\noindent

\begin{figure*}[!h]
\centering
\includegraphics[width=0.9\linewidth]{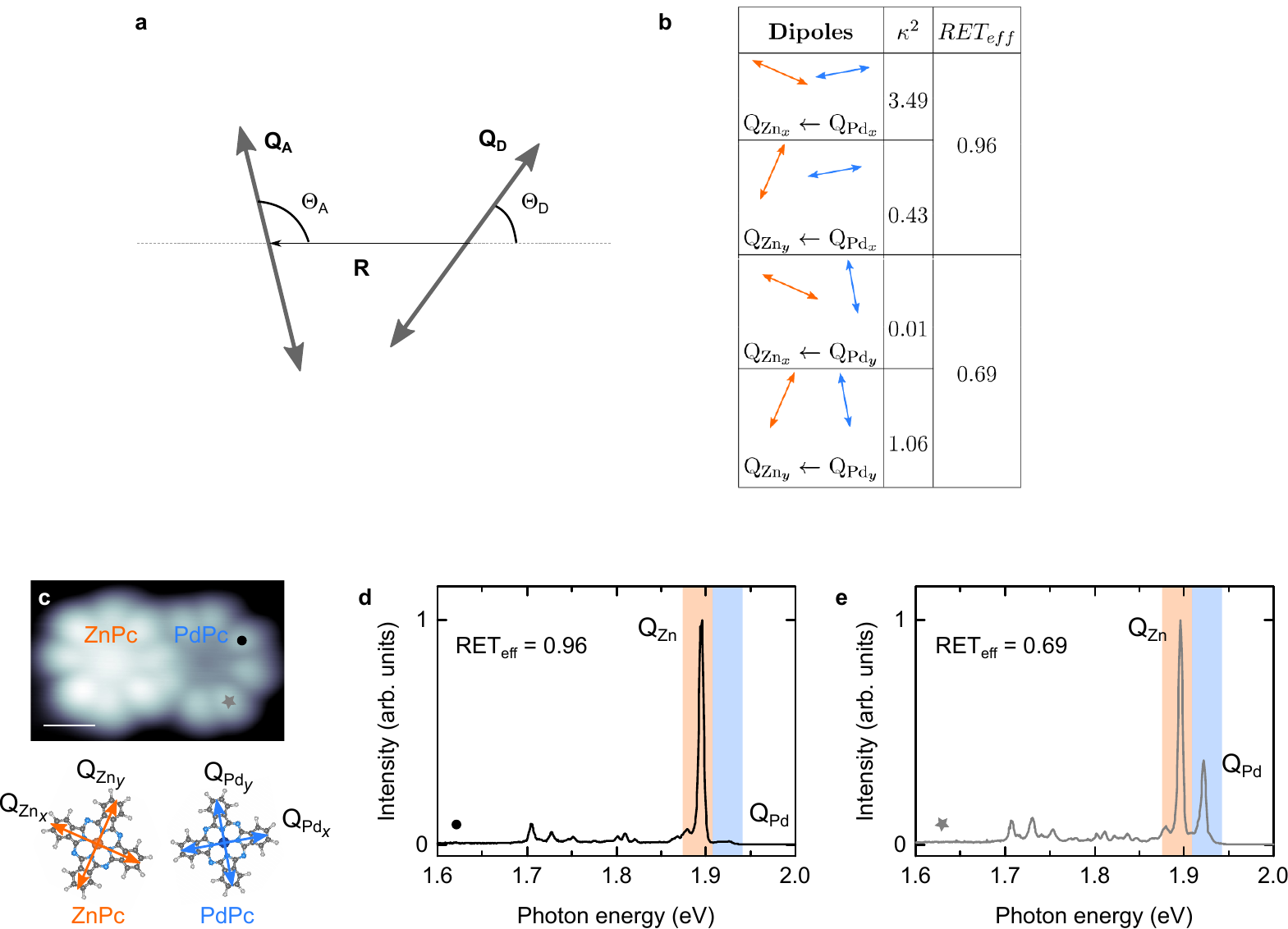}\caption{The effect of the relative dipole orientation on RET. (a) Geometrical configuration of the donor and acceptor transition dipoles in a dimer. \textbf{R} is the vector joining the centers of the dipoles. (b) $\kappa^{2}$ values calculated for the dimer configurations presented in (c). (c) Top: STM image ($V = -2.5$\,V, $I = 10$\,pA). Scale bar 1 nm. Bottom: ball-and-stick models of the PdPc--ZnPc donor--acceptor pair with the respective dipoles indicated. (d,e) Plasmon-corrected STML spectra ($V = -2.5$\,V, $I = 300$\,pA, acquisition time $t = 300$\,s) recorded at the positions marked by a black dot (d) and gray star (e) in (c). The main emission lines of each molecule are highlighted in color in the spectra. These data are also presented in Fig. 2a and Fig. 2d of the main manuscript.}
\label{figureS2}
\end{figure*}

\newpage
\noindent
\clearpage

\clearpage
\pagebreak

\begin{figure*}[!t]
\centering
\includegraphics[width=0.7\linewidth]{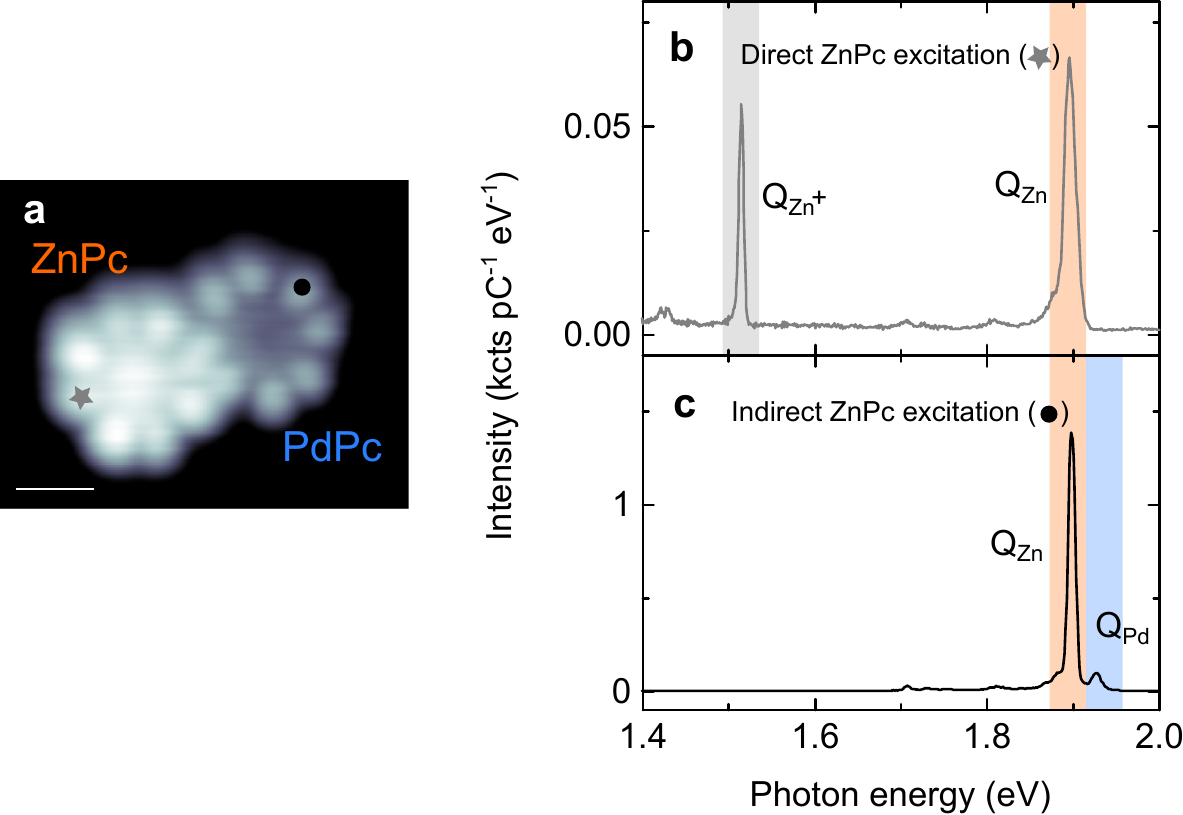}\caption{Charge state of the ZnPc molecule. (a) STM image ($V = -2.5$\,V, $I = 10$\,pA) of a PdPc-ZnPc dimer. Scale bar 1 nm.  (b) and (c) STML spectra recorded at positions marked in (a), $V = -2.7$\,V, $I = 300$\,pA, acquisition time $t= 180$\,s for (b) and $t = 100$\,s (c). Note the different vertical scales.}
\label{figcharge}
\end{figure*}

\clearpage
\newpage

\begin{figure*}[!t]
\centering
\includegraphics[width=0.8\linewidth]{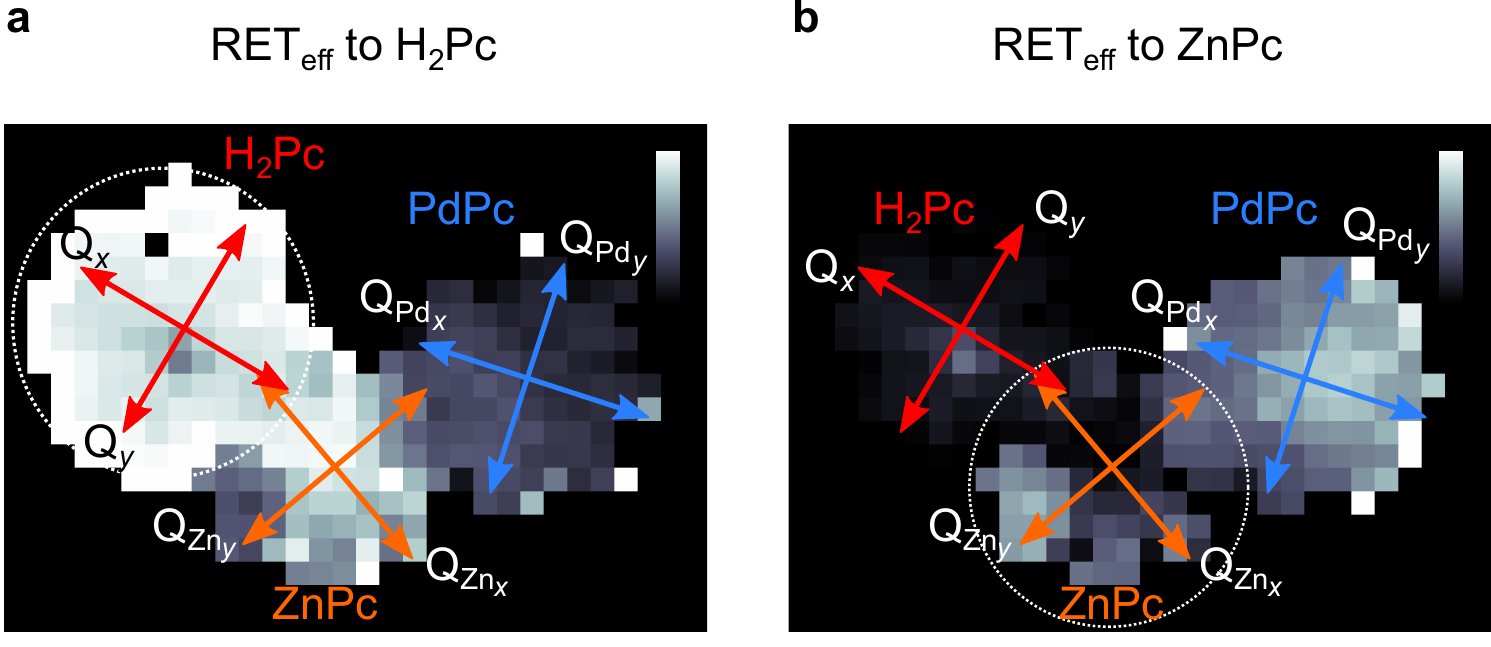}\caption{RET efficiency maps for H$_{2}$Pc and ZnPc as acceptors in the PdPc-ZnPc-H$_{2}$Pc trimer. The circles mark the spatial extension of the acceptor, the arrows indicate the transition dipoles of the labelled chromophores. The high-intensity areas denote precise locations where the sub-molecular excitation of the donor results in an efficient energy transfer to the acceptor (H$_{2}$Pc in (a) and ZnPc in (b)). The color scales range from 0 to 1.}
\label{figureS5}
\end{figure*}

\clearpage
\pagebreak

\begin{figure*}[!t]
\centering
\includegraphics[width=0.9\linewidth]{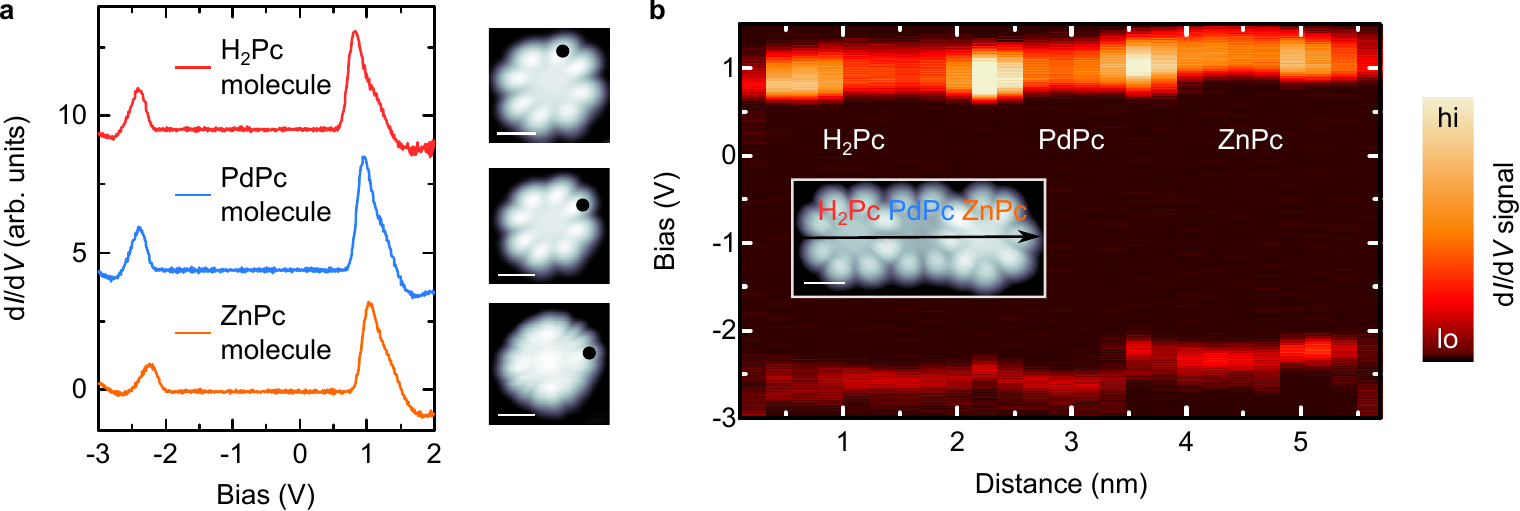}\caption{Comparison of the single-molecule and trimer d$I$/d$V$ spectra. (a) Left panel: d$I$/d$V$ spectra recorded on individual H$_{2}$Pc, PdPc and ZnPc. Set-point: $V$ = -3 V, $I$ = 15 pA. Right panel: STM images of the corresponding individual molecules, the dots indicate the positions where the d$I$/d$V$ spectra were acquired. $V$ = -2.5 V, $I$ = 5 pA. (b) A series of 25 d$I$/d$V$ spectra acquired along the H$_{2}$Pc--PdPc--ZnPc trimer (following the black arrow in inset). Set-point: $V$ = -3 V, $I$ = 15 pA. Inset: STM image of the studied trimer, $V$ = -2.5 V, $I$ = 5 pA. All scale bars are 1 nm.}
\label{figureS_didv}
\end{figure*}

\begin{figure*}[!t]
\centering
\includegraphics[width=0.9\linewidth]{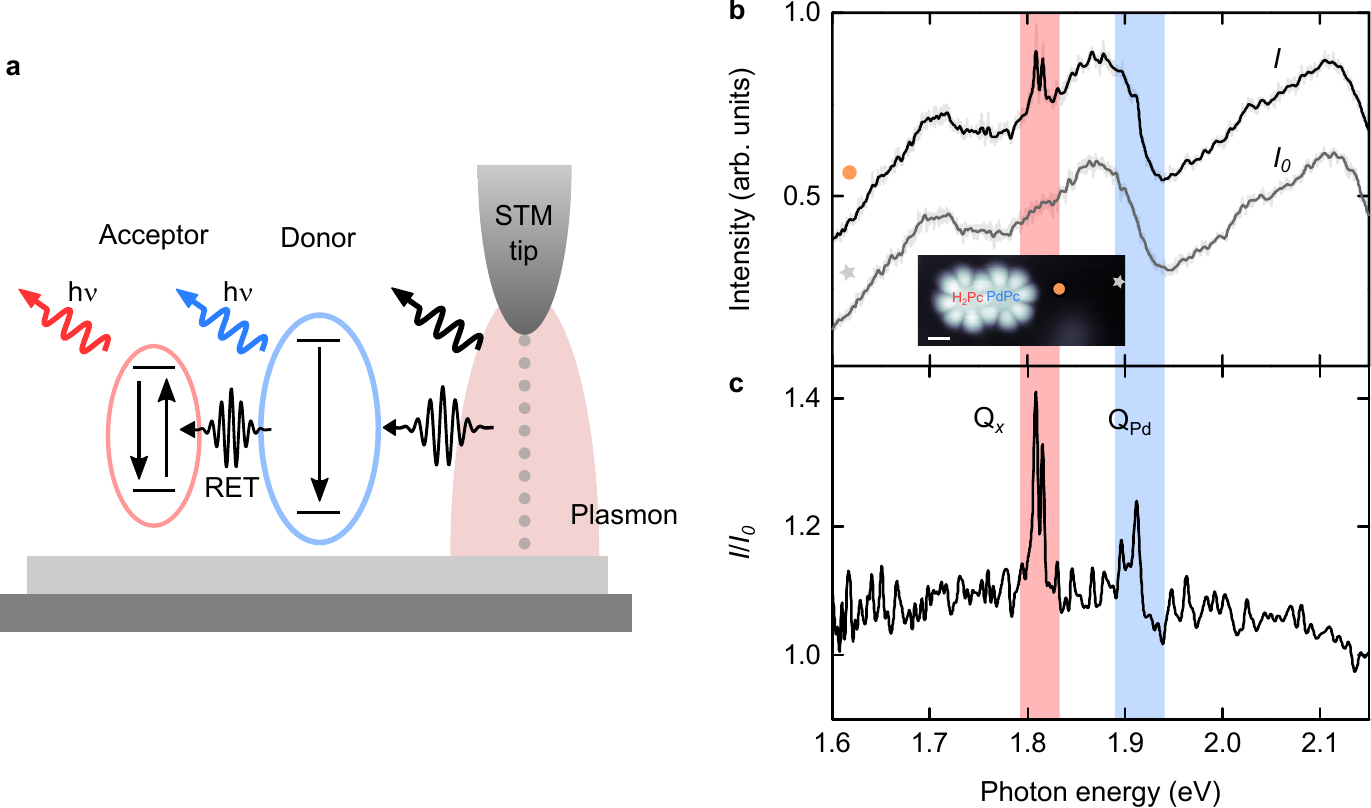}\caption{RET excited ''at-distance''. (a) Sketch of the experiment. (b) STML spectra acquired with the STM tip located at a distance of r = 2.2 nm (upper curve, $I$) and r = 4 nm (bottom curve, $I_{0}$)  from the center of the PdPc molecule. Inset: STM image $V$ = -2.5 V, $I$ = 5 pA of the investigated PdPc--H$_{2}$Pc dimer. The dot and star marked the positions at which the spectra have been recorded. Scale bar 1 nm. (c) Normalized spectrum $I/I_{0}$.}
\label{figureS_atdistance}
\end{figure*}

\begin{figure*}[!t]
\centering
\includegraphics[width=0.87\linewidth]{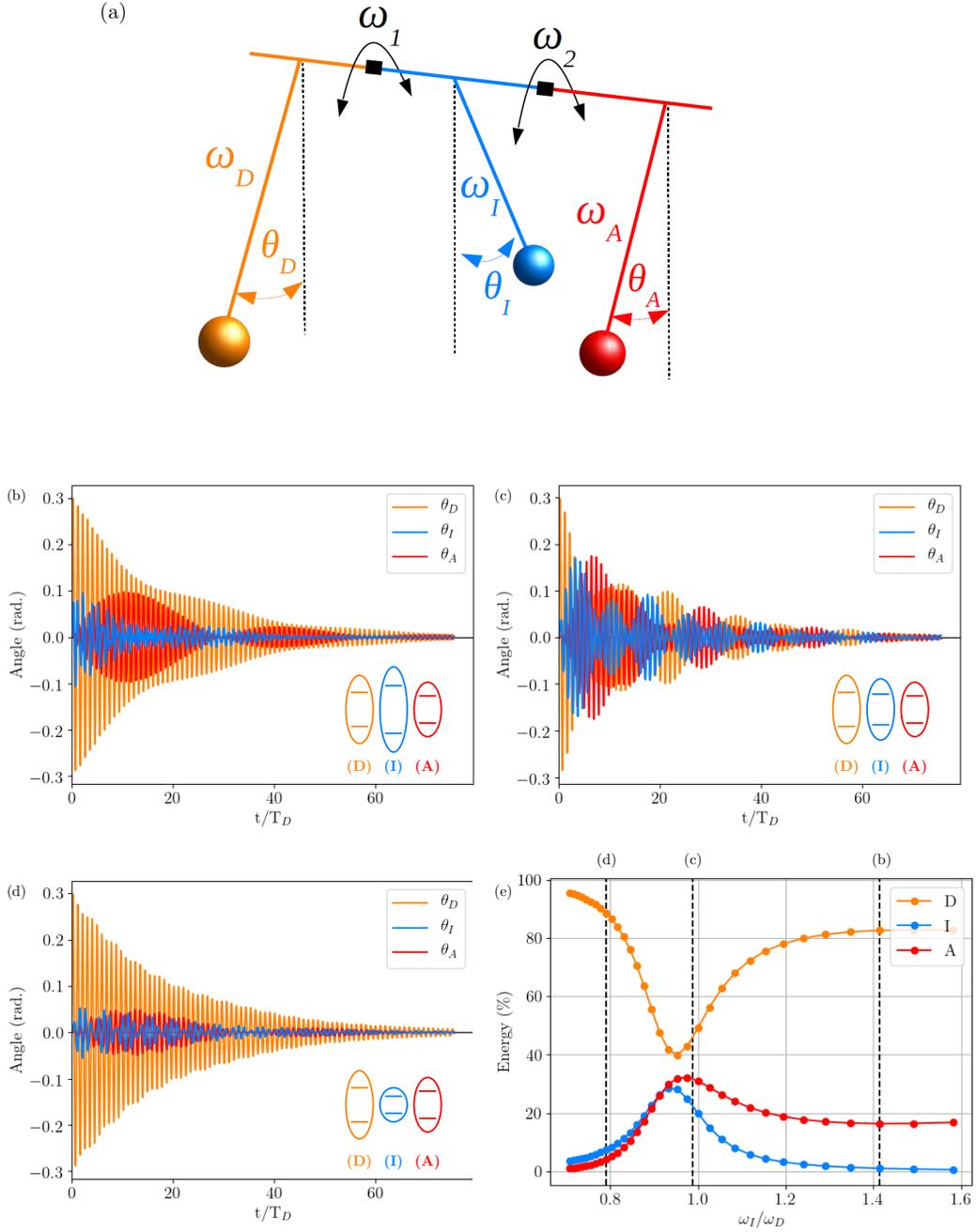}
\caption{Modeling the role of an intermediate molecule with a classical oscillatory approach. (a) Graphical representation of the three-pendulum model. Oscillation amplitudes of the three pendulums as function of the normalized time $t$/$T_{D}$ for a large (b), a medium (c) and a small (d) eigenfrequency of the intermediate pendulum, where $T_{D} = 2\pi/\omega_D$ . (e) Fraction of the excitation energy dissipated by the donor (D), the intermediate (I) and acceptor (A) pendulums as a function of the normalized intermediate pendulum eigenfrequency $\omega_I/\omega_D$.}
\label{figSI_model}
\end{figure*}

\bibliographystyle{naturemag}

\end{document}